# DIELECTRIC ROD FEED FOR COMPACT RANGE REFLECTOR

Balabukha N.P., PhD, Basharin A.A., PhD, and Shapkina N.E., PhD

Institute Theoretical and Applied Electromagnetics Russian Academy of Sciences,

ITPE RAS, 13 Izhorskaya, Moscow, Russia, 125412.

## Abstract

A dielectric rod feed with a special radiation pattern of a tabletop form used for the compact range reflector is developed and analyzed. Application of this feed increases the size of the compact range quiet zone generated by the reflector. The feed consists of the dielectric rod made of polystyrene; the rod is inserted into the circular waveguide with a corrugated flange. The waveguide is excited by the $H_{11}$-mode. The rod is covered by the textolite biconical bushing and has a fluoroplastic insert in the vicinity of the bushing. Mathematical modeling was used to obtain the parameters of the feed for the optimal tabletop form of the radiation pattern. The problem of the electromagnetic radiation was solved for metal-dielectric bodies of rotation by method of integral equations with further solving of the problem of the synthesis for feed parameters. The dielectric rod feed was fabricated for the X-frequency range. Feed amplitude and phase patterns were measured in the frequency range 8.2-12.5 GHz. Presented results of mathematical modeling and measurements for X-range radiation patterns correlate well. It is shown that this feed increases by 20-25% the quiet zone of the compact range with reflector in the form of nonsymmetrical cutting of the paraboloid of revolution 5.0 x 4.5 m in size in the frequency range 8.5-10.0 GHz as compared to a conical horn feed.

## 1. Introduction

With compact range collimator design, the quiet zone size with respect to a reflector size is of great importance, as in the quiet zone the amplitude-phase distribution of the field main polarization component should be nearly uniform. The problem of construction of collimator with a maximal ratio of the quiet zone square to the reflector square is topical.

One of the lines of the collimator quiet zone increase is the development of feeds with a specific form of the amplitude radiation pattern main lobe which is close to the tabletop form and with uniform phase radiation pattern. In this case in the central part of the reflector the signal is nearly constant by amplitude and phase. The signal level decreases in the vicinity of the collimator edge. It is required to obtain a low level of the feed radiation pattern in the reflector edge direction. In this case the reflector edge is weakly excited, then an amplitude of the edging diffracted wave and the nonuniformity of the amplitude-phase distribution of electromagnetic field in the quiet zone of the compact range decreases correspondingly.

## 2. Description

Considered rod feed is used in the collimator with a reflector in the form of a non-symmetric cutting of the paraboloid of revolution 4.5 m in height, 5.0 m in width and with a focal distance of 3.5 m. In order to obtain the quiet zone of the appropriate size the feed should possess the amplitude nonuniformity 1 dB and phase nonuniformity 3° in the angular sector ±21° with respect to the feed axis or the center of the quiet zone as well as amplitude decay less -15 dB in the reflector edge direction (±42° with respect to the feed axis). The dielectric rod feed with such radiation pattern is presented in Fig.1.

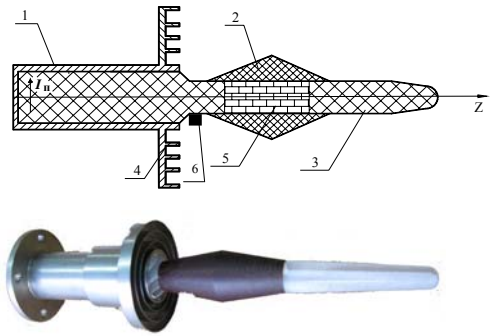

Fig.1. Dielectric rod feed with fluoroplastic insert.

The following properties of the electromagnetic field in the dielectric rod and near it [1] were taken into account when the construction of the dielectric rod feed was chosen: it is possible to consider that the field of the antenna is the field of the nonuniform plane wave propagating along the dielectric rod; field energy is transmitted by waves that propagate both in the rod and outside it; the ratio of the powers of these waves depends on the rod material dielectric permittivity ε and on the ratio of the rod diameter to the wavelength.

If the power ratio is close to unit, it is possible to obtain the tabletop radiation pattern which meets all named requirements by means of variation of the current phase on the definite part of the rod length. Polarization currents phase may be changed, for example, varying the diameter of the part of the rod, which jut out from the waveguide. It can be realized by disposition of the conic bushing on the rod, made of the material different from the rod material.

## 3. Results

Dielectric rod feed consists of the dielectric rod 3 made of polysterene (permittivity ε=2.5), dielectric biconical bushing 2 made of textolite (ε=4.5), corrugated flange 4 and circular waveguide 1 powered by the wave $H_{11}$. The dielectric rod has a fluoroplastic insert 5 (ε<2.0) in the vicinity of the biconical bushing (see Fig.1).

The radiation pattern of such feed is approximately equal to the sum of three components: the axial-symmetric radiation pattern of the dielectric rod with a maximum in the rod axis direction; the bushing radiation pattern of the funnelar shape with a minimum in the named axis direction; radiation pattern of the fluoroplastic insert with a maximum directed along the feed axis.

With optimal parameters of the rod, biconical bushing, and fluoroplastic insert and with meticulous focusing of all elements of the feed construction one can obtain nearly tabletop feed radiation pattern in E- and H-planes.

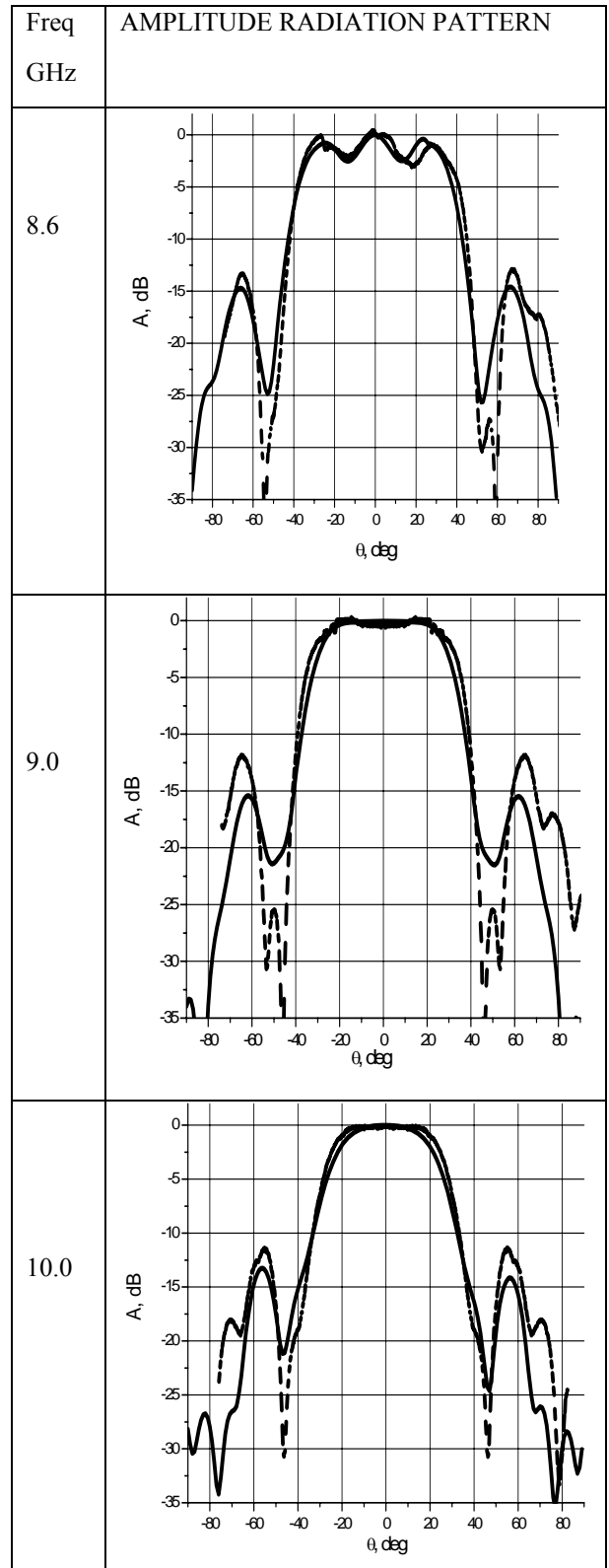

| Freq GHz | AMPLITUDE RADIATION PATTERN |
|---|---|
| 8.6 | |
| 9.0 | |
| 10.0 | |

Fig.2 a. Amplitude radiation pattern of the dielectric rod feed with fluoroplastic insert.

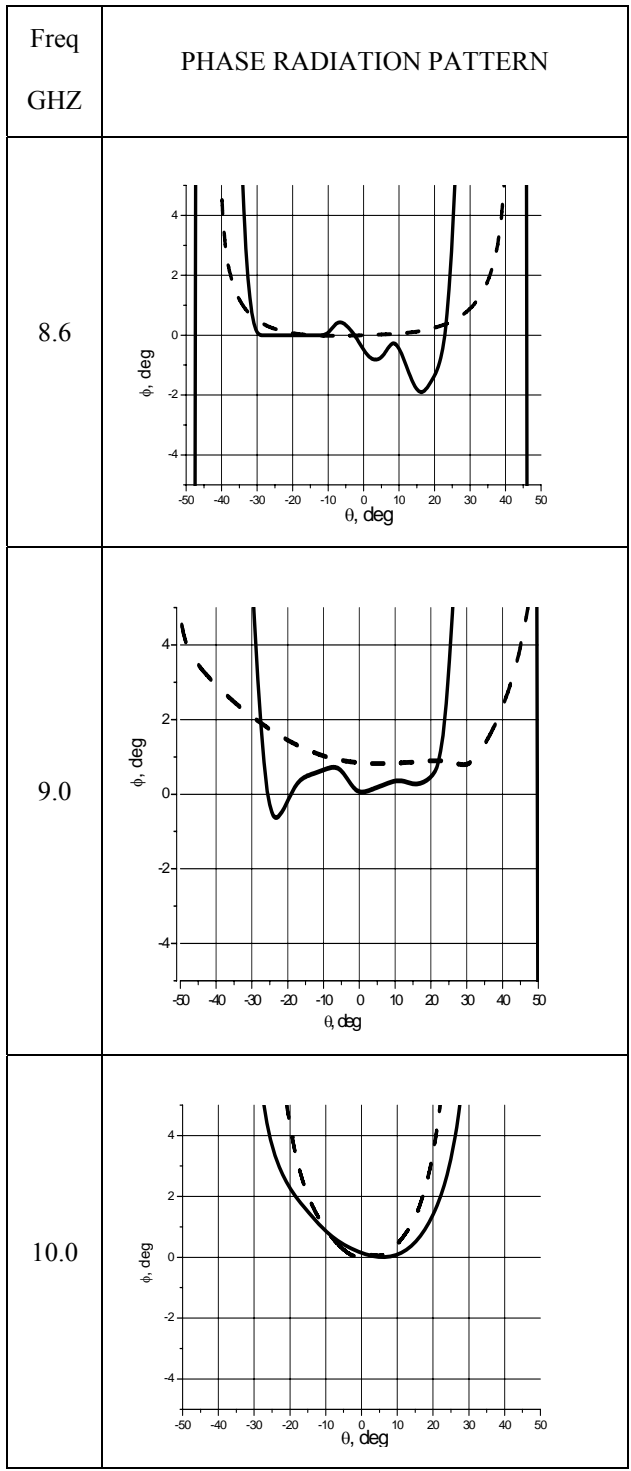

Fig.2 b. Phase radiation pattern of the dielectric rod feed with fluoroplastic insert.

To optimize the feed, numerical analysis of the feed radiation pattern was conducted by the method of integral equations for the bodies of rotation based on Green tensor functions.

The method was developed by E.N. Vasil'ev [2].

The symmetry of bodies of rotation permits to reduce the analyzed problem to the solution of the one-dimensional integral equation with respect to the currents on the surface of the analyzed feed taken as a metal-dielectric body of rotation.

In Fig. 2 numerical results (solid line) and measurement results (dashed line) are presented for amplitude (Fig. 2a) and phase (Fig. 2b) radiation patterns of the dielectric rod feed in the frequency range 8.6 - 10.2 GHz. Amplitude radiation pattern of the feed at low frequencies has three humps, with frequency increase they transform to two humps and then to one hump (cosine form). The phase radiation pattern is sufficiently uniform in the vicinity of the amplitude radiation pattern main lobe. In this case reflector edges are radiated by the electromagnetic wave with intensity less -20 dB with respect to the field intensity in the central part of reflector; thus the amplitude of the electromagnetic wave scattered by reflector edges decreases significantly. Presented calculation and measurement results correlate well in X-range.

In Fig. 3 the dependence of the nonuniformity (solid line) of the feed amplitude radiation pattern main lobe is presented in the angular sector $\pm21^{\circ}$ with respect to the axis of the dielectric rod feed with a fluoroplastic insert. In the same figure the corresponding dependence for the feed without an insert is presented (dashed line). As shown in graphs, amplitude nonuniformity less 1 dB is obtained in the frequency range 8.8—10.2 GHz for the dielectric rod feed with the fluoroplastic insert. The same result for the feed without an insert is obtained in the frequency range 9.0—9.4 GHz.

It was found that the phase center of the dielectric rod feed with the fluoroplastic insert is on the axis of the dielectric rod in the plane of cutting of the biconical bushing directed to the free end of the rod. The position of the phase center varies significantly with the frequency change; thus it provides significantly precise alignment of the phase center and reflector focus in the required frequency range.

In Fig. 4 results of the numerical modeling by method of physical optics [3] are presented for electromagnetic fields in the quiet zone of the compact range formed by the 5-meters reflector: in the central horizontal and vertical cross-sections of the quiet zone for the reflector radiation by the dielectric rod feed (solid line) and horn conical (dashed line) feed at the frequency 10.0 GHz.

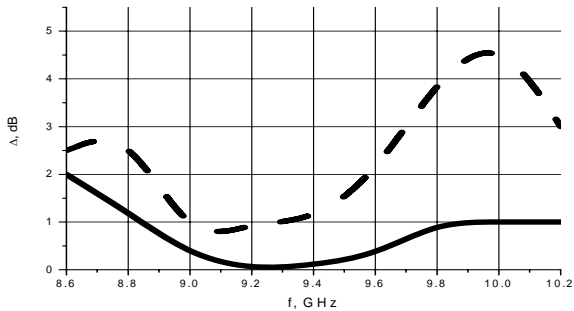

Fig. 3. Dependence of the nonuniformity of the feed amplitude radiation pattern main lobe in the angular sector ±21° with respect to the axis of dielectric rod feed with fluoroplastic insert (solid line) and without insert (dashed line).

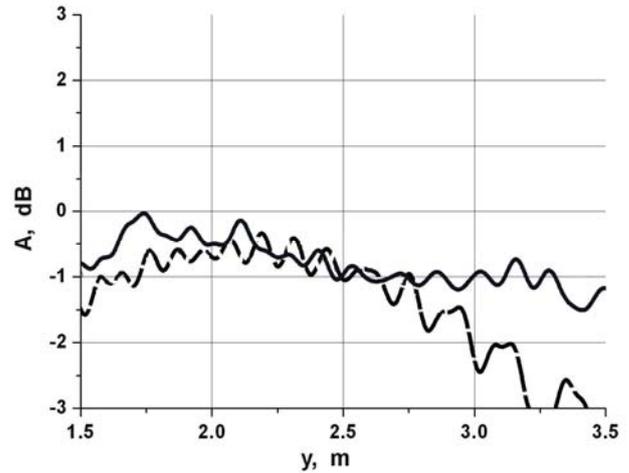

Fig.4b. Amplitude field distribution in the central vertical cross-section of the quiet zone for the reflector quiet zone.

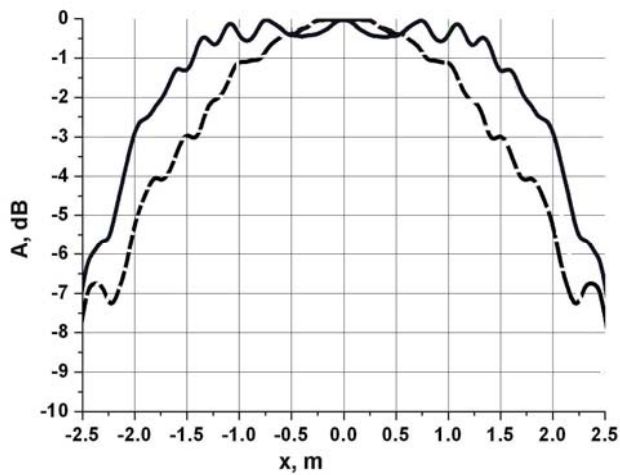

Fig.4a. Amplitude field distribution in the central horizontal cross-section of the quiet zone for the reflector quiet zone.

Similar field distributions are observed in the frequency range 8.8-10.5 GHz. In Fig. 4 amplitude field distributions in the central cross-section along the reflector focal axis in the horizontal plane are presented in Fig. 4a and in the vertical plane in Fig. 4b.

In offset reflectors there is a supplementary nonuniformity of the field amplitude in the quiet zone (Fig. 4) caused by the different distance between the feed and different points on the reflector surface as the reflector is nonsymmetrical with respect to the focal axis.

If that feed is used, the radiation pattern of which is symmetrical in the cross-section similar to the reflector symmetry plane, the nonuniformity is partly compensated by directing the feed radiation pattern maximum to the point on the reflector in the symmetry plane placed a bit further from the focal axis than the projection of the quiet zone center. The nonuniformity named above can be compensated more efficiently by using the feed with nonsymmetrical radiation pattern in the cross-section coincident to the reflector plane of symmetry. This radiation pattern is obtained if a metal semi-ring is set between the waveguide and conical bushing (Fig.1, pos. 5).

## 4. Summary

Application of the dielectric rod feed increases by 20-25% the quiet zone size of the compact range with reflector in the form of non-symmetric cutting (4.5 m x 5.0 m) of the paraboloid of revolution in the frequency range 8.5-10.0 GHz as compared to the conic horn feed.

## 5. References.